# Ultrasmall volume Plasmons - yet with complete retardation effects

Eyal Feigenbaum and Meir Orenstein[*]

*Department of Electrical Engineering, Technion, Haifa 32000, Israel*

## Abstract

Nano particle-plasmons are attributed to quasi-static oscillation with no wave propagation due to their subwavelength size. However, when located within a band-gap medium (even in air if the particle is small enough), the particle interfaces are acting as wave-mirrors, incurring small negative retardation. The latter when compensated by a respective (short) propagation within the particle substantiates a full-fledged resonator based on constructive interference. This unusual wave interference in the deep subwavelength regime (modal-volume<$0.001\lambda^3$) significantly enhances the Q-factor, e.g. 50 compared to the quasi-static limit of 5.5.

---

[*] Electronic address: meiro@ee.technion.ac.il

Optical cavities with extreme merits are explored intensively [1,2]. In the era of nanoscience – cavities supporting ultrasmall modal volumes are of great interest for applications such as sensing molecular-size substances [3], coupling to single "atom" emitters for single photon sources [4] or nano-lasers [5], compact memory cells [6], enhanced local nonlinearities [7] and more. Fundamental issues – related to the limits of photon localization far below the standard half-wavelength ($\lambda/2$) and strong light matter interactions are intriguing as well.

At the lower scale of photonics and plasmonics, reside the nano metalic particles acting presumably as quasi-static cavities or antennas and exhibiting large local field enhancement yet with low Q-factor [8,9] for visible and near-IR light. Here the energy storage is attributed to the localized plasma dipoles oscillations with no retardation (electromagnetic wave propagation) effects. It was recently shown that in this limit the specific structure of the nano-particle is not important and the Q-factor is bounded (from above) by the material properties namely $Q=(\omega\partial_\omega\varepsilon_M')/(2\varepsilon_M'')$ ($\varepsilon_M=\varepsilon_M'+i\varepsilon_M''$ is the metal relative dielectric constant [9] and $\omega$ is the angular frequency). For Gold nano particles, even at a relatively long wavelength of 1.5μm, the Q-factor is only ~5. The authors of Ref. [9] are wondering whether retardation effects can improve the quasi-static plasmonic cavity, especially its low Q factor.

At the larger scale, photonic and plasmonic cavities are based on wave retardation effects – namely on constructive wave interference, where the energy is stored mainly at the enhanced field intensity. Very high Q-factors are reported, but the cavity mode size is always larger than half the wavelength ($\lambda/2$). Realization of nano- cavities based on interference is facilitated by the fact that plasmon-polaritons – mixed electromagnetic and plasma density waves that propagate along metallic surfaces, may exhibit significantly reduced wavelength (slow-wave). By employing these slow plasmon-polaritons, some nano-cavities were reported having modal volume $V=(36nm)^3=0.006\cdot\lambda^3$ with Q=170 (calculated) for a whispering gallery mode cavity in a plasmonic gap [10]; V=(186nm)$^3$=0.015·$\lambda^3$ with Q~30 (calculated) for a plasmonic Fabry Perot cavity [11]; V=(86nm)$^3$=0.006·$\lambda^3$ with Q=10 (measured) for a loaded plasmonic Fabry Perot [12].

The theme of this letter is related to a possible role of such wave propagation effects in determining the cavity characteristics of the nano metalic particle discussed earlier, although the latter's dimensions are much smaller even compared to a plasmon-polariton wavelength. It is of a basic interest to comprehend why and when wave interference may have a major importance in ultra small structures – where light (or polaritons) cannot accumulate almost any phase in propagation. Furthermore the existence of such retardation effects and their influence on the cavity characteristics may change the quantitative interpretation of many phenomena based on nano particle cavities – such as Surface Raman Enhancements [13], nano-shells [7] as well as on related design rules (e.g. nano-plasmonic lumped circuitry [14]).

We show that such retardation effects can enhance the particle Q-factor significantly, compared to the quasi-static value, which results in remarkable Q/V figure of merit, namely enabling enhanced strong light matter interactions. The underlying effect is wave reflection from 'band-gap' mirrors (to be explained later) surrounding the particle, which exhibits small but negative phase – and together with the short wavelength characterizing the plasmon-polariton yields the dramatic effect.

It is easy to convey the basic physical mechanism by examining the classical parallel plate Fabry-Perot resonator - storing light energy by constructive interference of waves, based on round- trip phase accumulation of $2\pi m$ ($m$=0,1,2,…). The overall accumulated phase stems from the combination of light propagation and the reflection phase from the cavity mirror interfaces. In regular cavities the two mirrors contribute either $2\cdot'0'$ or $2\cdot'\pi'$ phase (mirrors with respective lower or higher refractive index compared to that of the cavity), thus not participating effectively in determining the cavity resonance condition. Consequently, constructive interference is achieved by setting the cavity length to $m \cdot (\lambda/2)$ and the smallest optical cavities are $\lambda/2$ long (e.g. in VCSELs, and in moderate Q-factor defect cavity). However, special mirrors can exhibit a amplitude reflectivity with small negative phase ($-\delta$), such that the two cavity mirrors contribute either $2 \cdot ('0'-\delta) = -2\delta$ or $2 \cdot ('\pi'-\delta) = 2\pi - 2\delta$. In both cases a very short propagation in the cavity segment is enough to complement the $2\pi m$ (m=0 or 1 respectively) round trip phase, namely – just the distance required to overcome the phase progress in the mirrors segments ($-2\delta$). Such a negative phase delay in reflection can be easily derived from the complex Fresnel equation (as depicted in Fig. 1(a)), which yields the following mathematical condition: for $|n_C| > |n_M|$ (cavity index>mirror index) the condition is $n_{Cr} n_{Mi} < n_{Mr} n_{Ci}$ (r and i are the real and imaginary parts respectively) and for $|n_C| < |n_M|$ the condition is the inverse of the above. This can be realized by using lossy media – but we seek for a solution that has a reflection coefficient magnitude of ~1, which is essential for achieving a reasonable Q-factor for such ultra-short cavities. Thus the favorable solution is employing a lossless mirror that supports a decaying field – namely a 'band-gap mirror'.

As an example an air cavity is constructed between two regular silver plated 'shaving mirrors'. If the metal mirrors were perfect conductors, the minimum cavity length for light at 1.5μm will be $\lambda/2$ = 750nm. However, the metal (Ag – parameters taken from Palik [15]) is rather a plasma and our wavelength is within the band-gap (below the plasma frequency) which results in an evanescent field in the metal and a reflection phase (of the two mirrors) $2 \cdot ('\pi'-0.07\pi) = 2\pi - 0.14\pi$ (see Fig. 1(b)). The smallest is only 52nm long (0.07×$\lambda$/2) where the wave propagation complements the missing phase of 0.14π. The quality factor of such a cavity is determined by the Ag conduction losses (no radiation loss in this configuration) and is Q~6, according to the term [16]:

$$Q = \frac{\int \left( \eta |H|^2 + \partial_\omega (\omega \varepsilon') |E|^2 \right) dv}{\int \left( \partial_\omega (\omega \varepsilon'') |E|^2 \right) dv} \quad (1)$$

where $\eta$ is the wave impedance, H and E are the magnetic and electric fields respectively. A related example, in the microwave regime, is addressing a specially designed metamaterial interface, which was used to reduce a resonant antenna size, employing also a negative phase [17-19]. It should be noted that the simple Fabry Perot device of this section is also a first approximation to a nano plasmonic cavity formed between two very close-by Ag particles (a plasmonic dimer) to be mentioned later.

The same concept is applied now to a full fledged 3D particle-plasmon. We consider a cylindrical shape nano metallic particle cavity. The quasi-static Q-factor of such a cylindrical cavity – with all dimensions much smaller than half the wavelength, is determined by the complex metal dielectric constant as in [9]. However if we choose a specific nano-cylinder height, within this seemingly quasi-static regime, such that the phase accumulation due to the very short propagation along the cylinder axis will complement the reflection phases of the waves from the cylinder bases (due to the mismatch of metal and air), we may expect a retardation based cavity, as discussed for the 1D case above, with potentially enhanced Q-factor. It should be emphasized that the air surrounding the particle can be considered as a 'band-gap' mirror since very small particles virtually do not radiate and thus the field in the particle is coupled mainly to decaying near field in air. We analyzed a variation of this scheme – by embedding the nano cylinder in a coaxial cylindrical plasmonic shielding envelope (Fig. 2). This is done for several reasons: the structure eliminates completely radiation losses; the structure is more easily analyzed in closed form to elucidate the basic mechanisms; the outer shielding is another control parameter on the mode volume; such a configuration is amenable to applications such as nano probing and nano fluidics.

The cavity is analyzed as a short plasmonic coaxial line segment, coupled on both sides to a plasmonic hollow cylindrical waveguide of the same diameter (Fig. 2). The lowest order plasmonic mode ($TM_0$) of the coaxial segment does not exhibit a cut-off. Therefore, for small radii the only propagating mode is the $TM_0$ within the coaxial segment ('resonator') while all other modes including the $TM_0$ of the waveguide segments ('mirror segments') are evanescent. Here the 'band-gap mirrors' with a negative phase delay are realized by a modal cutoff, which is very similar in nature to the plasma band-gap of the previous section. It should be noted that when considering the metal losses – all modes are complex, but we keep the original naming which is justified by the relatively small losses of our configuration.

A rigorous solution is employing all modes of the two plasmonic waveguide segments (coaxial line and hollow waveguide). The azimuthal symmetry of the structure results in negligible coupling of the $TM_0$ mode to TE and TM modes of higher azimuthal orders. The dispersive metal permittivity ($\varepsilon_M$) is fitted to experimental values [15] according to the complex Drude model (a very plausible assertion for $\lambda_0=1.5\mu m$) ($\varepsilon_M=-86.1-i8.16$, $\partial_\omega(\omega\varepsilon_M)=86.6+i16.2$). Applying transfer matrix calculations based on the orthogonal eigen-modes of the resonator (1) and the mirror segment (2) we obtain the reflection coefficient by:

$$\underline{\rho} = \left[\underline{\underline{AB}}+\underline{\underline{I}}\right]^{-1}\left[\underline{\underline{AB}}-\underline{\underline{I}}\right]\underline{\iota} \qquad A_{m,n} = \frac{\left\langle e_n^{(2)}\middle|h_m^{(1)}\right\rangle}{\left\langle e_m^{(1)}\middle|h_m^{(1)}\right\rangle} \qquad B_{n,m} = \left(\frac{\left\langle e_n^{(2)}\middle|h_m^{(1)}\right\rangle}{\left\langle e_n^{(2)}\middle|h_n^{(2)}\right\rangle}\right)^* \qquad (2)$$

where $e$ is radial E-component and $h$ is azimuthal H-component. The vectors $\iota$ and $\rho$ are the impinging and reflected amplitudes of the cavity mode. The major reflection amplitude is of the propagating $TM_0$ field reflected to the counter-propagating $TM_0$ field while negligible power is carried by the higher modes (evanescent) in the coaxial segment. Indeed – such a single mode approximation yields quiet accurate results – as can be seen by comparing the continuous and dashed red lines of Fig. 3(b), showing that the main effect is due to the decaying field of the zero-order mode in the mirror segment. For an accurate calculation of the reflection phase, virtually all modes on both of the facet's sides were considered, showing a good convergence to the boundary conditions with less than 5 evanescent modes on both sides for the diameters of interest. The results for the shortest cavity ($L_0$) are depicted in Fig. 3(a) exhibiting a substantially shorter cavity than the *effective* $\lambda/2$, which is due to the negative phase accumulation in the mirror. In contrary to the Fabry-Perot with plasma mirrors of the previous section – the mirror segment (mostly air here) is the higher impedance segment, such that the reflection phase is slightly below '0' (Fig. 3(b)) rather than the below '$\pi$' value of the previous case. The resonance condition is fulfilled for $(m=0)\cdot\pi/2$ where the mirror interfaces are contributing negative phase and the short propagation complements the phase to '0'.

The Q-factor is calculated directly from the energy distribution and in Fig. 4(a) we show a range of parameters where the Q-factor surpasses the quasi-static value ($Q_{static}$~5 is indicated by a dashed line in the Figure). As the size of the cavity is enhanced the Q-factor as well as the modal volume (V) are enhanced (modal volume is calculated from the intensity variance and since the field negligibly penetrates to the metal nano cylinder, the modal volume is almost excluding the nano particle volume). The figure of merit Q/V (Fig. 4(b)) exhibits very high values – in the range of $10^6$-$10^8$, and is enhanced monotonically as the resonator dimensions are reduced. Fig. 5 depicts a typical nano-cavity modal intensity distribution at resonance with very small volume (a=30nm, b=90nm, L=60nm). Even though the intensity is mostly localized around the inner metal interface, the shield plays a significant role in limiting the radial as well as the longitudinal dimensions of the mode, by inducing a smaller effective wavelength (slower plasmonic wave). This effect contributes both to a faster phase accumulation in the resonator and closer-to-zero reflectance phase (also apparent in the preceding figures). The resulting modal volume of this example is strictly nano-scaled, having average effective dimension of 75nm ($V=0.0002 \cdot \lambda^3$), an order of magnitude smaller than the effective wavelength of the SPP mode (~750nm at $\lambda_0=1.5\mu m$). The Q-factor is 30, ~6-times higher than the maximum value expected from quasi-statics. As the transversal radii of the 3D structure (*b* and *a*) are reduced a faster decay into the band-gap mirror is experienced and the reflection phase is further approaching '0' ( Fig. 3(b)) – resulting in shorter cavities.

The potential impact of this retardation related effect on nano particle-plasmons is of importance for many typical scenarios in the field. A signature of this effect is exemplified by calculating (full numerical finite element) the Q-factor of a plasmonic dimer cavity– (two almost touching gold particles). The Q-factor calculated under the electrostatic approximation is ~6 (similar to the shape-independent result of [9]), while the wave solution results in ~11 – for a randomly selected (not wave-resonant) inter-particle spacing (Fig. 6). At the relatively low Q-factor regime of these cavities, the wave-retardation is important not only exactly at resonant dimensions but also off resonance – resulting in larger energy storage than the values perceived today.

We have shown that ultrasmall particle-plasmon exhibits a significant signature of propagating plasmon-polaritons, exploiting a nearly zero but yet finite negative wave- reflection phase on the facets. The proposed concept was exemplified for simplified 1D structures and subsequently studied with full fledge analysis for a realistic 3D plasmonic cavity, exhibiting $V \sim 10^{-4} \cdot \lambda^3$ with respectful Q-factor of 30. It is important to note that the specific nano-cavity discussed here– can be actually fabricated either vertically or horizontally by metal and dielectric layer deposition and subsequent nano-etching by focused ion beams or by e-beam patterning - the resulting cylindrical nano- metallic particle will be embedded in a dielectric e.g. $SiO_2$, and surrounded by metal shielding, while light can be input or extracted via the finite length waveguide mirror segment by tunneling. The in/out coupling should be made smaller than the metal losses to preserve the Q-factor value. Such a cavity may be applied for variety of strong interactionsfor sensing and basic physics studies.

**Figures**

Figure 1

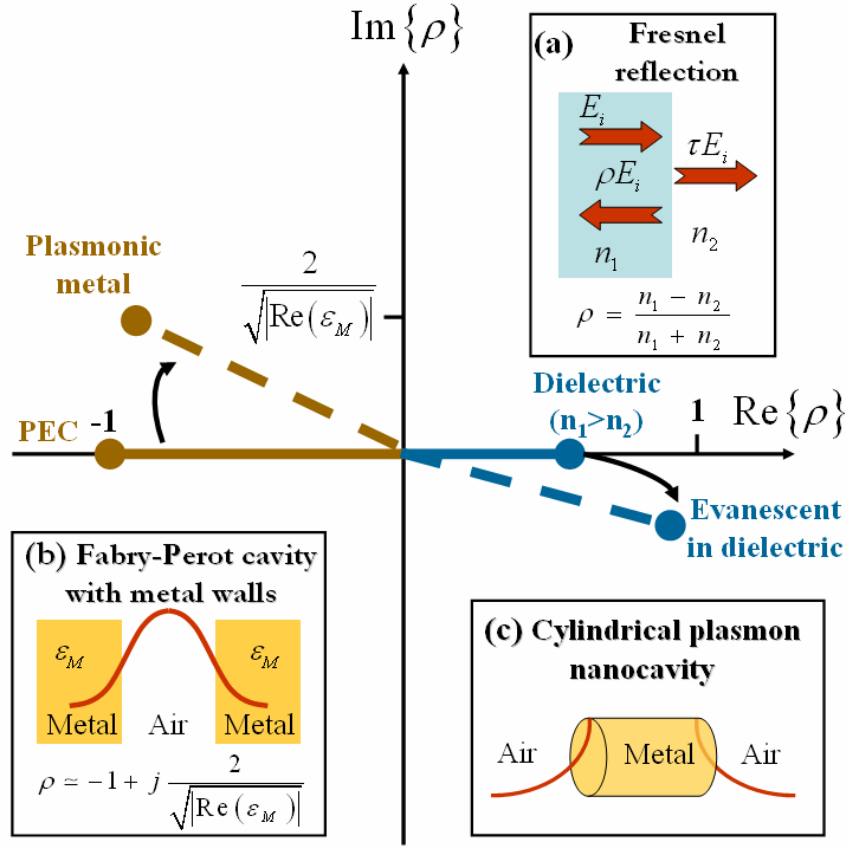

FIG. 1. The complex (normal incidence) reflection coefficient (ρ) at various cases. The insets are: (a) Fresnel equation for plane-waves reflection coefficient; (b) metal-air-metal based plane-waves cavity, showing phase progress relatively to PEC mirror, due to the plasmonic mirrors (less than 'π' phase); (c) The 3D nanoparticle cavity and the related reflection phase progress (less than '0' phase)

Figure 2

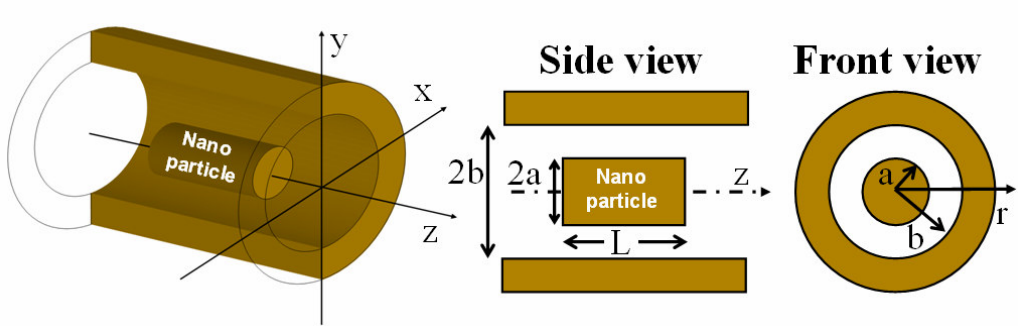

FIG. 2. A 3-dimentional illustration, of the particle-plasmon shielded cavity. The structure consists of an inner nano-cylindrical particle and longer (coaxial) outer shielding layer, both made of plasmonic metal (Gold).

Figure 3

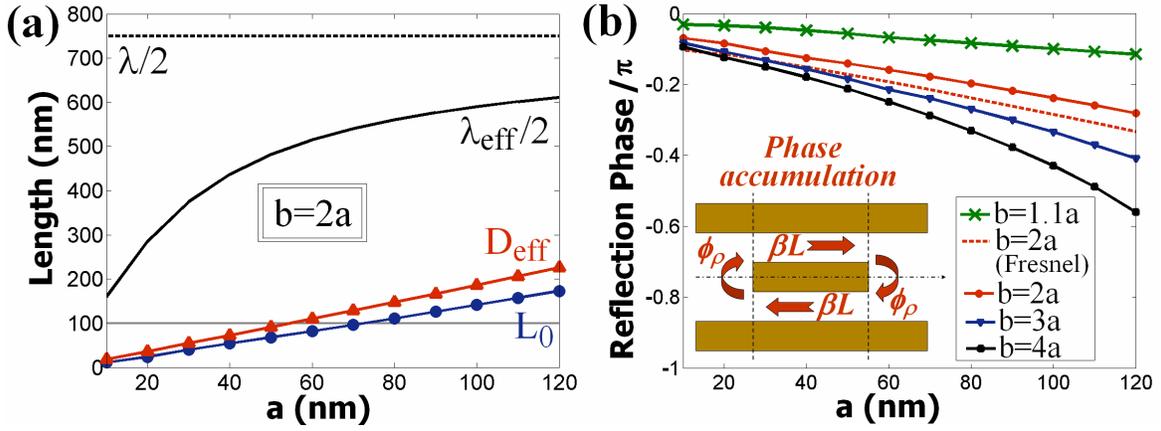

FIG. 3. Cavity characteristics as a function of the nano particle radius ($\lambda_0=1.5\mu m$). (a) The various characteristic length scales: blue - shortest cavity length, red - effective modal dimension and black - half the plasmonic wavelength in the resonator segment; (b) Reflection phase ($\phi_\rho$). Dashed red line - Fresnel-like approximation using the effective index of the lowest mode ($TM_0$) in all segments for b=2a. Inset: illustration of the phase accumulation cycle.

Figure 4

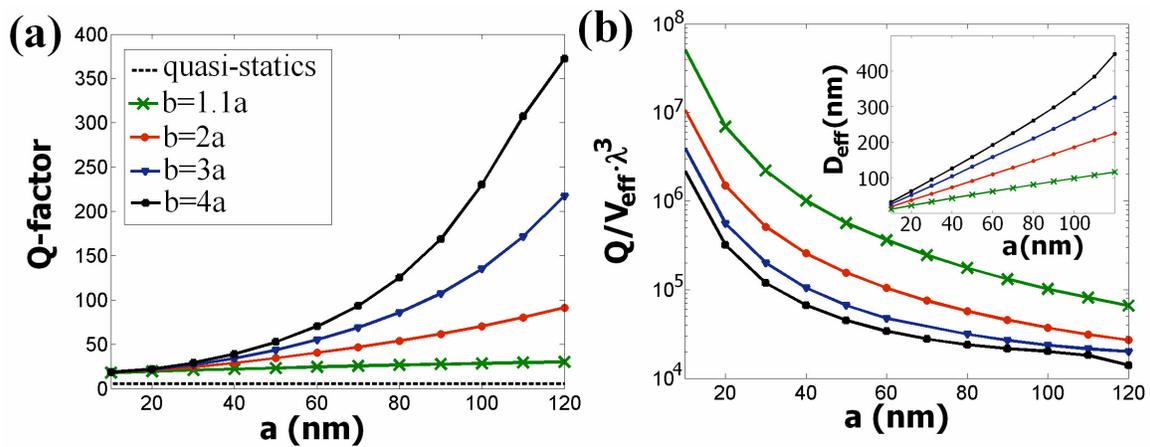

FIG. 4. Cavity merits for as a function of the nano particle radius ($\lambda_0=1.5\mu m$). (a) Q-factor values. Black dashed line: Quasi-static value (~5) (b) Corresponding Q/V figure of merit. Inset: the effective modal dimension: $V^{1/3}$.

Figure 5

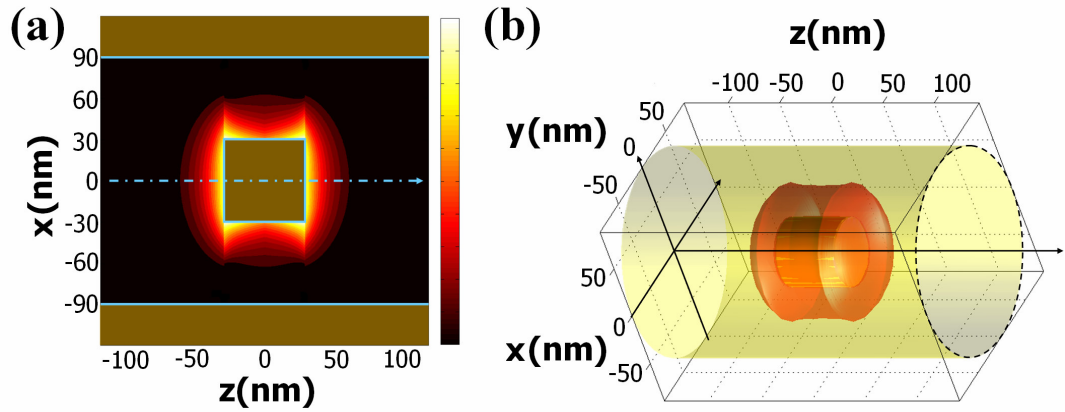

FIG. 6. Field intensity distribution and its cross-section for a typical Gold & air cavity mode: a=30nm, b=90nm, L=60nm, Q=30 ($Q_{statics}$=5.5), V=(75nm)$^3$, $\lambda_0$=1.5μm.

Figure 6

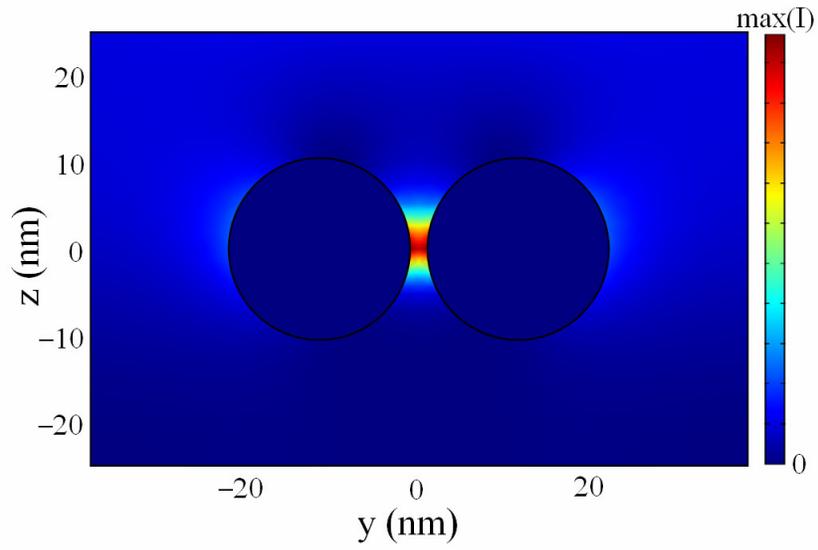

FIG. 6. Field intensity of (2D) Au ($\varepsilon_M$=-86.1-i8.16) dimer excited by a y-polarized plane wave. Diameter is 20nm, and inter-center distance is 23nm. Q quasi-static is~6 and Q full wave is ~11.